# Growth Stress Induced Tunability of Dielectric Constant in Thin Films

*K. V. L. V. Narayanachari,[†] Hareesh Chandrasekar,[‡] Amiya Banerjee,[‡] K. B. R. Varma,[†] Rajeev Ranjan,[§] Navakanta Bhat,[2] and Srinivasan Raghavan [*,†,‡]*

[†]Materials Research Centre, [‡]Centre for Nano Science and Engineering, [§]Materials Engineering, Indian Institute of Science, Bangalore 560012, India

## ABSTRACT

It is demonstrated here that growth stress has a substantial effect on the dielectric constant of $ZrO_2$ thin films. The correct combination of parameters – phase, texture and stress – is shown to yield films with high dielectric constant and best reported equivalent oxide thickness of 0.8 nm. The stress effect on dielectric constant is twofold, firstly, by the effect on phase transitions and secondly by the effect on interatomic distances. We discuss and explain the physical mechanisms involved in the interplay between the stress, phase changes and the dielectric constant in detail.

*Corresponding author e-mail: sraghavan@cense.iisc.ernet.in

The scaling of silicon-based CMOS technology has hit a limit in terms of the thickness of $SiO_2$ based dielectric schemes,[1] which are now limited by quantum tunneling at values less than 2 nm and carrier mobility in the Si channel.[2] The first problem has been overcome by the use of high-k dielectrics,[3,4] like $ZrO_2$ and $HfO_2$ among many others.[5-8] The issue of low carrier mobility necessitates the use of materials with high electron and hole mobilities[9] and development of suitable dielectrics with sub-1 nm equivalent oxide thickness (EOT) for these new channel materials. The most promising channel material is Ge having electron and hole mobilities of 3000 and 900 $cm^2$/Vs respectively.[10] Dielectric wise, $ZrO_2$ is known to form a stable interface with Ge [5,11,12] and hence $ZrO_2$/Ge is a potential combination that can replace the conventional $SiO_2$/Si pair for CMOS applications.[13,14]

The electrical properties of any metal oxide semiconductor (MOS) device are critically dependent on the dielectric-semiconductor interface trap density ($D_{it}$), permittivity (k) of the dielectric and leakage current from the gate through the dielectric. For dielectrics on Ge, the problems of obtaining a low interface trap density and a high value of permittivity are now tackled independently of each other.[15] The $D_{it}$ can be reduced by various treatments such as S-passivation, epi-Si, $GeO_2$ interlayer, forming gas anneal, UV ozone oxidation,[16] Ar Gas anneal,[17] and/or surface passivation before high-k deposition.[15,18,19] In the present study we hence concentrate on tailoring the dielectric permittivity of $ZrO_2$ films obtained by reactive sputtering. In a previous paper[4] we had documented the effect of sputtering parameters, especially growth rate, on texture and stress selection in $ZrO_2$ films on Si. In this work we show that stress in the films can have a significant effect on the dielectric permittivity of $ZrO_2$ films on Ge. The correct choice leads to sub-1 nm EOT with acceptable leakage currents.[20]

Low resistivity n-Ge <100> was used as the substrate for $ZrO_2$ film deposition. A series of pure-$ZrO_2$ films on germanium were grown by off-axis reactive direct current (R-DC) sputtering in an ultra-high vacuum (UHV) chamber with base pressure of $4\times10^{-9}$ Torr. The UHV chamber geometry and details of film growth using R-DC are published elsewhere.[4] Metal-oxide-semiconductor (MOS) capacitor structures were made using shadow mask deposition of platinum dots onto $ZrO_2$. Real time stress measurements during thin film growth were performed using a curvature based stress measurement technique using a multiple beam optical stress sensor (MOSS).[21] The MOSS is installed at normal incidence and curvature measurements are performed on an unclamped substrate facing up.[4,22,23]

The key dielectric results to be discussed in this work are shown in Figure 1. It summarizes the effect of thickness at two different growth rates, 0.6 and 4.2 nm/min, on the dielectric constant of $ZrO_2$ films deposited on Ge substrates. Given our previous results on the effect of growth rate on film microstructure[4]—dense films are grown at rates <5 nm/min whereas those grown at >15 nm/min are porous—all the films in present study were grown at <5 nm/min to get compact films with smooth surfaces (RMS roughness < 1 nm).[20] Three points are noteworthy. Firstly, in all cases the dielectric constant increases with thickness. Secondly, the dielectric constants measured for some of the zirconia films are so low that one could qualify them as low-k dielectrics. Thirdly, for films deposited at 0.6 nm/min, the dielectric constant at comparable thickness is about an order of magnitude higher than in films grown at 4.2 nm/min. The permittivity of 23 measured in the 5 nm film, (first data point in Figure 1) deposited at 0.6 nm/min results in an EOT of 8 Å, which is less than the best EOT of 1.6 nm achieved in sputtered $ZrO_2$ by Nagai *et al*.[24] and comparable to the best reported $ZrO_2$ EOT of 5–8 Å by Chui *et al*.[25] using the UV ozone technique. It must be noted that there is no effect of temperature or annealing in forming gas on the dielectric constant of high growth rate films. This third point is the key result of this paper. The reason for this behavior is now discussed in terms of the stress, texture and phase composition of the deposited films.

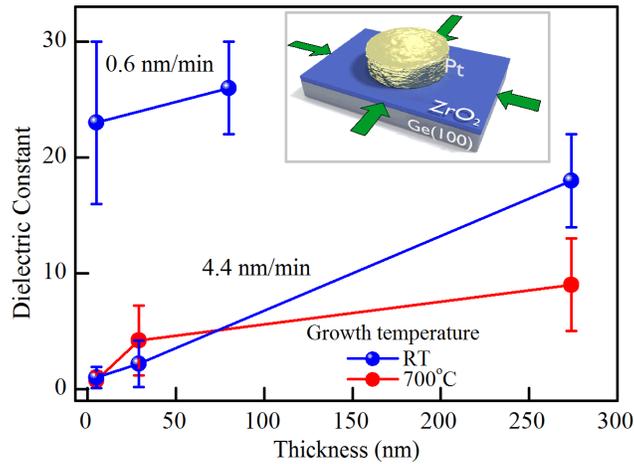

Figure 1. Dielectric constant vs. thickness for films grown at two growth rates, error bars were calculated from at least three samples for each data point. Inset shows the MOS stack schematic used, wherein substrate and Pt dot were the bottom and top contacts for electrical measurements. The arrows indicate that the stress state in the film is equibiaxial.

Bulk pure $ZrO_2$ exists in three different forms - monoclinic from room temperature to 1170°C, tetragonal until 2370°C and transforms to cubic thereafter. The orientationally averaged dielectric constants of the amorphous, monoclinic, cubic and tetragonal phases are experimentally measured to be 22, 20, 37 and 47 respectively.[26,27] In addition, for the tetragonal and monoclinic phases, as calculated from first-principles,[26 20] the anisotropic lattice dielectric constant along different directions can vary from as low as 16.7 along the c-axis of the monoclinic phase to as high as 46.6 along the a-axis of the tetragonal phase. This illustrates the importance of phase selection and orientation in modifying the dielectric constant. In unconstrained systems such as powders, the high temperature phases can be stabilized at room temperature by a reduction in size due to surface energy effects. Since, these phase transformations involve a change in volume, phase stability is also dependent on strain in constrained systems. Hence, in the case of thin films, such as the ones used in this study, both size and strain effects on phase stability are expected to be present which affect the measured dielectric constant.[16,28] It should be pointed out that strain can change the dielectric constant of a given phase by itself directly or by engendering a phase transformation.

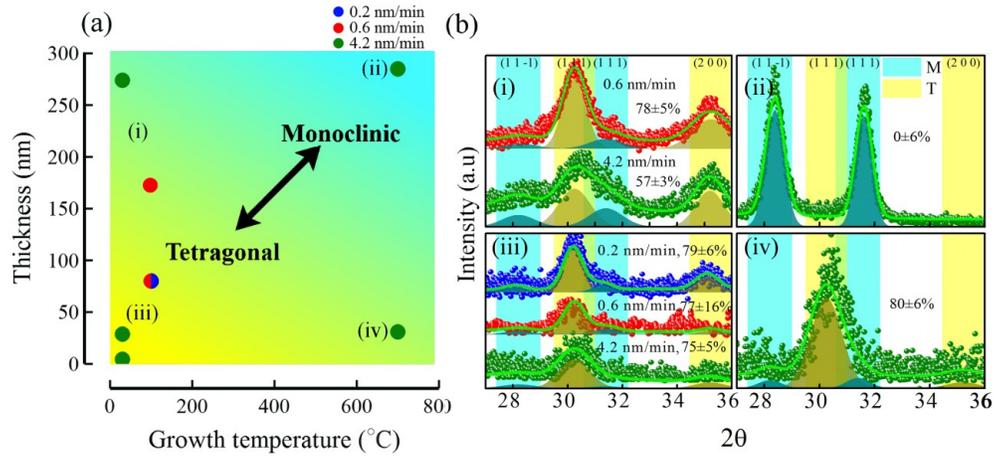

**Figure 2.** (a) Map summarizes the effect of temperature, thickness and growth rate on the phase composition in the sputtered ZrO$_2$ films. (b) XRD patterns corresponding to various points in Figure 2 (a). Tetragonal (or cubic) phase fractions obtained from the areas under the peaks are mentioned on the graph.[29] The remaining is monoclinic.

Given the importance of phase and its orientation on the dielectric constant, the phase composition of these films as determined by x-ray diffraction is summarized in Figure 2. In brief, all the zirconia films deposited as part of this study are polycrystalline mixtures of the tetragonal (or cubic) and monoclinic phases. For the very thin films used in the current study it is difficult to differentiate between the tetragonal and cubic phases of zirconia. As listed in

, and shown in Figure 2(a) the crystallite size and monoclinic phase content increase with thickness and growth temperature. For the thick (>250 nm) films in Figure 1, the increased monoclinic content is the reason for the lower dielectric constant of films grown at higher temperatures in comparison to those grown at room temperature ((i) vs (ii) in Figure 2(b)). For the thin films, <100 nm, (See Figure 2(b)(iii)), the x-ray diffraction patterns however, apparently, show that there is very little effect of growth rate on the phase composition. Thus, the large jump in dielectric constant seen on reducing the growth rate from 4.2 to 0.6 nm/min, it appears, cannot be explained by phase composition alone. It must be pointed out that noise in the XRD spectra for very thin films makes accurate estimation of phase fractions difficult. While, the issue of the absolute value of the dielectric constant being so low for the 4.2 nm/min-thin films will be discussed later, the films deposited at 0.6 nm/min have larger dielectric constants than even the thicker films deposited at 4.2 nm/min. The same also applies to the 4.2 nm/min-thin film grown at high temperatures (See Figure 2(b)(iv)) which in spite of being predominantly tetragonal has a very low dielectric constant. Thus, in summary, phase composition alone cannot explain the dielectric trends in Figure 1. As discussed in what follows, growth rate dependent stress forms the third axis of the plot in Figure 2(a) and helps to complete the story.

**Table. 1.** Growth rate and film thickness dependent crystallite size in nanometers.

|  | Growth temperature | |
|---|---|---|
|  | **25°C** | **700°C** |
| **5 nm** | *2.5* | *3* |
| **~25 nm** | *5* | *7* |
| **~250 nm** | *8* | *12* |

(left side label: **Thickness**)

In our previous work we had shown that growth rate has a significant effect on the in plane equibiaxial stress of such zirconia films. In brief, we had demonstrated that changing the growth rate from ~0.1 nm/min to 5 nm/min can change the growth stress from -3 GPa to +0.2 GPa respectively. This results from the fact that at low growth rates adatoms are injected into the grain boundaries in the growing polycrystalline film resulting in compressive stress generation.[30] As growth rate increases, the magnitude of such injection decreases and so does the compressive stress. Eventually, at very large rates only stresses due to boundary formation, which are tensile in nature remain, yielding films with a tensile stress.

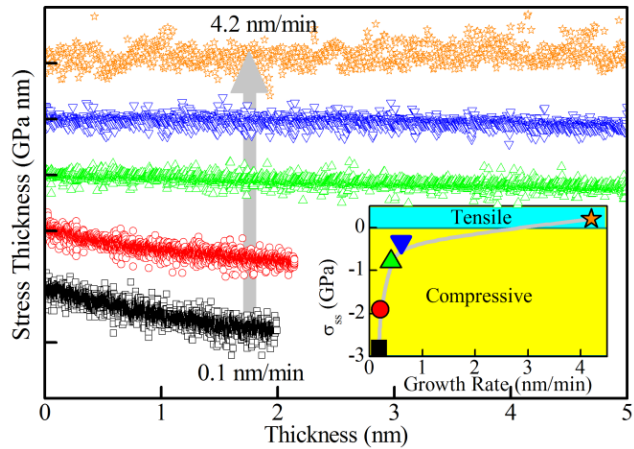

**Figure 3.** Plot showing stress thickness data for films grown at different growth rates. Arrow indicates the direction of increasing growth rate. On the y-axis each division corresponds to 10 GPa nm. Inset shows film steady state stress values extracted from the slope of the stress thickness plots and the growth rate dependency.

The stress thickness vs. thickness plots of films deposited at different rates, (0.1–4.2 nm/min) are shown in Figure 3 and the effect of growth rate on stress is summarized in the inset. It can be very clearly seen that a decrease in growth stress from 4.2 nm/sec to 0.6 nm/sec results in a change in the in plane compressive stress from -3 to +0.2 GPa. The reader is reminded that an in-plane compressive stress would result in out of plane tensile Poisson strains and vice versa. An in-plane compressive stress is also expected to stabilize the tetragonal phase while a tensile stress the monoclinic phase.

In order to determine if stress is the reason for the effect of growth rate on dielectric constant, in addition to the samples already discussed, zirconia films with different room temperature stress values were hence synthesized by depositing them at different rates and at different temperatures. Both thick, >80 nm, and thin <10 nm films were grown. The stress-dielectric constant dependence of these films is summarized in Figure 4. It is essentially the 3-dimension of the plot in Figure 2 (a). The stress plotted on the x-axis is the film stress at room temperature. For films grown at room temperature there is only a growth stress component. For those grown at elevated temperatures the thermal expansion mismatch stress has been added to the measured growth stress (see supplementary material).[20] It is clearly seen that for both categories of films the dielectric constant peaks at a mildly compressive stress of about -0.3 GPa before dropping down to lower values on either side. Even though the three factors enumerated before phase composition, texture and stress can affect the dielectric constant along the film normal, the data, as will be discussed, shows that in the films of the present study, the effect is entirely stress driven either directly or indirectly (through the stress-phase interdependence). Thus, stress tuning is critical to obtaining the highest possible dielectric constants. For purposes of discussion we divide this graph into a direct stress effect, the left half and an indirect stress effect, the right half.

The reduction in dielectric constant from the peak value with increasing in-plane tension is attributed to increasing fractions of the monoclinic phase. Even though this is not very obvious from the XRD data in Figure 2b(iii) due to noise levels in the data from these very thin films, it is well known from thin film zirconia literature that this indeed is the case.[28,31] This can also be very clearly seen from XRD data on thick films, Figure 2(b) (ii). The films with the highest dielectric constant (>25) are predominantly made up of a polycrystalline tetragonal phase (Figure 2(b) (i)) whose fraction is of the order of 77±16%, while those with a low dielectric constant are entirely monoclinic (Figure 2(b) (ii)). Even though they were grown at different temperatures, the crystallites sizes seen from are not too different and very much lower than the normally accepted 30 nm Garvie limit for the monoclinic to tetragonal transition in stress free unconstrained powders.[31,32] Thus, the transition to the monoclinic phase is due to the increasing biaxial tensile stress that favors the tetragonal to monoclinic diffusionless phase transition. This is the indirect stress effect.

The reduction in dielectric constant with increasing in-plane compression is however due to the direct effect of stress. As can be seen from the x-ray patterns in Figure 2, all of these films, grown at low rates, are tetragonal in phase composition. This reduction in dielectric constant is due to the fact that when films are under lateral biaxial compression, the lattice parameter along the film normal ($a_\perp$) increases due the Poisson effect. Since, the internal electric field in any dielectric material has been inversely proportional to the lattice constant ($E_i \propto a^{-3}$),[33] it results in reduced dielectric permitivity ($k_\perp$) measured along the same direction. A similar reduction in k with stress has been reported in literature - for $Ba_{0.5}Sr_{0.5}TiO_3$ and $SrTiO_3$.[20,34,35]

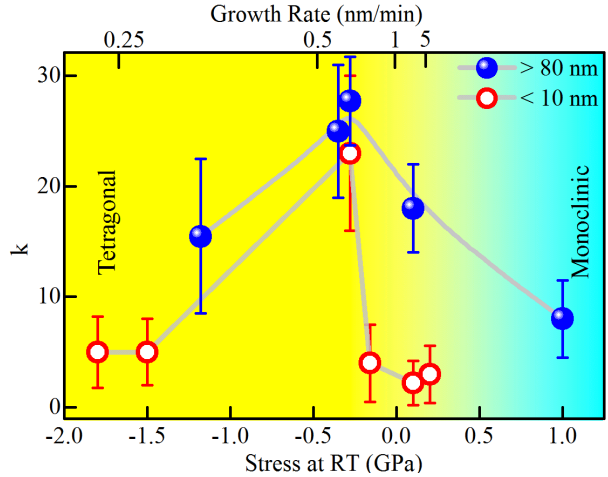

**Figure 4.** The effect of stress on the dielectric constant in $ZrO_2$/Ge films for two different sets of thicknesses.

The behaviour of the dielectric constant vs stress data seems to resemble the sharp increase in dielectric constant associated with paraelectric-ferroelectric phase boundaries in other oxide systems.[36] In addition, in the case of zirconia-hafnia systems the presence of a non-centrosymmetric orthorhombic phase with high dielectric constants of the order of 40 has been reported in literature.[37] This orthorhombic phase like electric behaviour was detected in pure zirconia films only at electric fields higher than 2 MV/cm. An orthorhombic phase was not detected in the pure zirconia films used in this study and the fields applied for the CV measurements were lower than 0.5 MV/cm. Thus, the presence of an orthorhombic phase and its effects is highly unlikely.

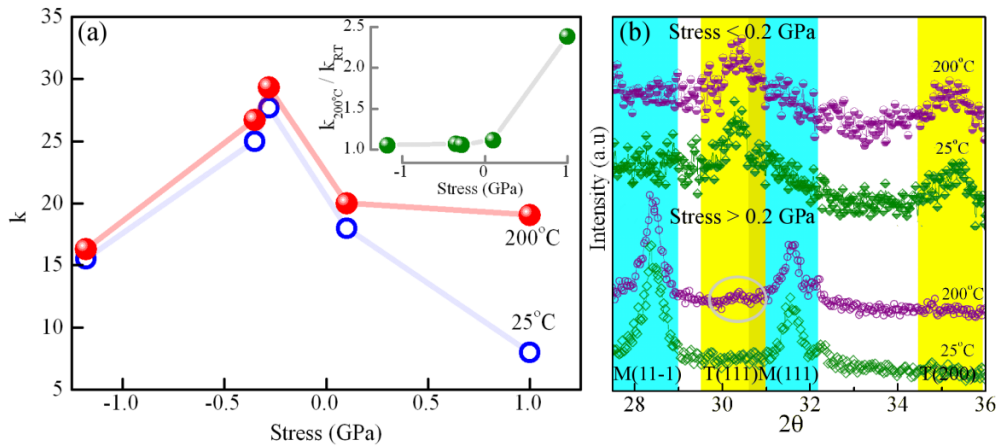

**Figure 5.** (a) Temperature dependent dielectric constant in thin films (80 nm) at room temperature and 200°C. Inset shows the relative change in dielectric constant. (b) X-ray diffraction patterns at room temperature and 200°C for the films with stress < 0.2 GPa and > 0.2 GPa.

In order to see if the reason for the peak of dielectric constant as seen in Figure 4 is due to phase instability—either of the structural tetragonal-monoclinic kind or of the paraelectric-ferroelectric kind—the films were cycled in temperature up to 200°C. As described in detail in the supplementary material, even though the increase in temperature showed a sharp increase in currents, most of this could be attributed to leakage and not due to an increase in the dielectric constant. The dielectric constant extracted after performing the relevant circuit analysis (plotted in Figure 5(a)) shows very little change between the 25°C and 200°C values. The only increase in dielectric constant, from 8 to 19, on increasing temperature was seen in the sample that was monoclinic at room temperature. This increase is accompanied by the appearance of a small tetragonal peak at 200°C as shown in Figure 5(b). This appearance of the tetragonal phase is due to both a rise in temperature as well as an increase in compressive stress due to thermal expansion mismatch between the film and substrate on heating. Cooling down resulted in a reversal to the same lower k values and same phase composition. Further, it should be noted (supplemental material) that the leakage current through the sub-1nm EOT dielectric is well within the leakage current specification for logic technology prescribed by ITRS.

Finally, we address the very low value of dielectric constants observed in some of the thin films deposited. In the case of the direct effect, based on the trends exhibited by even the thick films it appears that an increase in in-plane compression and therefore out of plane expansion is responsible for the very low values measured in the tetragonal phase. On the other hand the indirect stress effect (right half of the plot the data in Figure 3 and 4) seems to suggest that k is very sensitive to very small amounts (to the extent that they can be estimated by x-ray diffraction) of the tetragonal and monoclinic phases. For instance, in Figure 5 the appearance of a very small amount of the tetragonal phase results in a doubling of the dielectric constant. Thus, the very low k values in the thin films are presumably due to small fractions of a monoclinic phase. As shown in the supplementary material [26] cross sectional TEM micrographs reveal that the interface is clean and devoid of low-k oxides of Ge and therefore such oxides are not responsible for the low k observed.

In conclusion, controlled sputter deposition of $ZrO_2$ thin films on Ge was carried out for high-k applications. Growth conditions and thickness of the $ZrO_2$ films determine the stress levels and hence the phase content of the films. Films with an optimum value of stress exhibit the highest dielectric constant. Those with higher tension and high compression on either side of this optimum value have lower k-values due to either an indirect stress induced phase transition or a direct effect on interatomic distances. Films with an EOT of as low as 8 Å, among the best reported in the literature, and acceptable leakage currents have been demonstrated.[20]

## ACKNOWLEDGEMENTS

The authors acknowledge the support of the Department of Science and Technology (DST), National Nanofabrication Center (NNFC) and the Micro and Nano Characterization Facility (MNCF) at IISc, Bangalore, funded by MCIT and DST, Government of India. Narayanachari is thankful to CSIR for providing a senior research fellowship (SRF).